*Research Article*

# Gastroscopic Panoramic View: Application to Automatic Polyps Detection under Gastroscopy

Shi Chenfei [1,] Yan Xue [1,] Chuan Jiang [1,] Hui Tian [1,] Bei Liu [1,]
[1] *Women's Hospital School of Medicine, ZheJiang University, Hangzhou 310000, China*
Correspondence should be addressed to 21315079@zju.edu.cn

*Abstract*—Endoscopic diagnosis is an important means for gastric polyp detection. In this paper, a panoramic image of gastroscopy is developed, which can display the inner surface of the stomach intuitively and comprehensively. Moreover, the proposed automatic detection solution can help doctors locate the polyps automatically, and reduce missed diagnosis. The main contributions of this paper are: firstly, a gastroscopic panorama reconstruction method is developed. The reconstruction does not require additional hardware devices, and can solve the problem of texture dislocation and illumination imbalance properly; secondly, an end-to-end multi-object detection for gastroscopic panorama is trained based on deep learning framework. Compared with traditional solutions, the automatic polyp detection system can locate all polyps in the inner wall of stomach in real time and assist doctors to find the lesions. Thirdly, the system was evaluated in the Affiliated Hospital of Zhejiang University. The results show that the average error of the panorama is less than 2 mm, the accuracy of the polyp detection is 95%, and the recall rate is 99%. In addition, the research roadmap of this paper has guiding significance for endoscopy-assisted detection of other human soft cavities.

## 1. Introduction

Gastroscopy plays a major clinical role in the diagnosis of gastric diseases. The detection and diagnosis for gastric polyps by gastroscopic intervention is the most routine solution [1]. However, conventional endoscope diagnosis for polyps is prone to misdiagnosis, for the following reasons: first of all, as soft cavity, stomach is easy to deformation. Additionally, there're lots of folds in gastric inner wall, which leads to the result that gastric polyps aren't intuitive. During the examination, doctors need to move the camera lens of the endoscope back and forth to find polyps. What's more, doctors control the endoscope in vitro. Due to the narrow entrance and narrow vision, it's difficult for doctors to manipulate the lens flexibly to obtain a detailed and comprehensive observation of the gastric inner wall [2]. Last but not least, the following-up examination for polyps detection often relies on the initial ink injection in the previous procedure (see Figure 1). In this case, the ink injection area may fall off as time goes on or is dissolved by gastric mucosa, which result in missing the located polyps [3]. Based on the above reasons, it is certain clinical meaning to improve the method that detects and identifies the polyp lesion during the gastroscopic examination, to reduce the rate of misdiagnosis.

Many literatures have been proposed to assist gastroenterologists to detect gastric polyp and reduce the rate of misdiagnosis with modern science and technology [4]. For instance, some researchers have focused on the method that combines computer vision technology and conventional endoscope diagnosis to detect gastric polyps. In those researches, a typical example is that [5] proposed a method to estimate confidence distribution of polyps based on the polar matrix and covariance matrix. In this method, they used covariance matrix to prejudge the possible lesions in endoscopic image to assist doctors in diagnosis. However, the experimental result shows the detection accuracy depends on the range of viewing angles as camera moves, which limits its application. Besides, Gao et al [6] design a non-invasive biopsy mark system applied in gastroscopic examination. In this technology, they constructs a virtual static three-dimensional model of the gastric inner wall, by the means of CT. And the model can assist in intraoperative navigation and biopsy resetting, but the detection and accuracy of navigation is not perfect because it's difficult to accurately calculate the flexible deformation of the stomach by the static preoperative model. To this end, some researchers developed a set real-time three-dimensional reconstructed method for soft cavity, based on RGB image of endoscope diagnosis. The latter method solved the problem that in the preoperative model it's unable to calibrate dynamically the deformation of soft organs [7]. However, because the endoscope image texture feature isn't easily to be extracted, it's still an unsolved question that how to build stable and real-time models. In addition to three-dimensional models for the auxiliary examination of lesions, direct diagnosis in two-dimensional images is also an important research direction. For instance, [8] have proposed a new real-time method of pathological localization, in which they regard stomach peristalsis as regional affine transformation. However, in soft cavity organs, regional affine hypothesis is generally not valid. Furthermore, taking advantages of the non-invasive property, probe-based confocal laser endoscope (PCLE) are developed for assisting in quick detecting and positioning of polyps in real time. However, this technology relies on extra hardware solution and can't return visit and review the lesions [9]. There is another research direction of lesion examination which uses panorama technology to expend the inner wall of soft organs. Through the expended panorama, doctors can give a comprehensive and quick diagnosis for the inner wall of target organs, without the need to repeatedly check and change the viewing angle. It can avoid misdiagnosis caused by occlusion and other issues and meanwhile improve the diagnostic effect. However, panorama technology has high requirement for image splicing and fusing. It is prone to appear shadow, blur, or even dislocation between the spliced images. Besides, regional distortion of images in expended panorama can also affect the result of diagnosis [10-11].

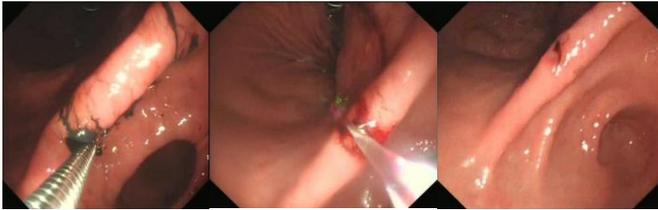

FIGURE 1: Polyps detection in traditional procedure. The lesions are determined by ink injection, but the ink may fade away before second-examination. And reflective areas can be found on the captured images.

Based on the above researches, this paper proposes an automatic diagnosis system for gastric polyps on basis of the panorama image of gastric inner wall. Specifically, the main innovative work of this paper shows as the following aspects: firstly, we build a panorama model for the gastric inner wall. Compared with the previous work, we don't rely on hardware to estimate camera position, so it's more available for operation. Furthermore, we optimize the method by means of optical consistency, and solve problems as registration error, blur and others caused by image stitching. Secondly, we develop an end-to-end panoramic multi-target detection network for gastroscopic procedure. Compared to conventional deep learning target detection framework, this paper uses the panorama image as network input, support the multi-target detection of polyps, and avoid distortion caused by stitching, which may mislead the doctors' diagnosis. Thirdly, we conduct clinical trials on entire system in affiliated hospital of Zhejiang University. The experimental result shows the model's error of our system is less than 2 mm and recalling rate of polyp detection is close to 100%. Our developed system can assist doctors in diagnosis and it helps reduce the rate of misdiagnosis, improve the efficiency of diagnosis and relieve the storage pressure of server data. Finally, the research method of this paper has a certain guiding significance for auxiliary diagnosis for other human soft organs, theoretically. In [23], a promising polyps detection method with CNN was also proposed. Compared with [23], the main significance of our method is our CNN framework considers the constructed gastric panoramic data as input. Moreover, our framework is designed as multi-target detection, which indicates our method can detect all the polyps of a patient just from one image.

The algorithm flow chart is shown in Figure 2. We take original image sequence from endoscope as the input of algorithm. After the image registration and the optimized texture fusion based on optical consistency theory, we obtain a more comprehensive view of the gastric inner wall. Then a proposed deep learning framework is worked on the generated panorama data and achieve automatic detection of gastric polyps.

Relative to conventional computer vision problems, there're lots of challenges to reconstruct panorama image of the gastric inner wall in the human soft cavity. These challenges can be summarized as the following aspects: first, it is hard to extract and match features, endoscope is usually a fisheye lens, so the images captured are always seriously distortion (see Figure 3). Furthermore, in soft cavity such as stomach, the inner walls are almost covered with mucosa, so the captured RGB images are prone to generate certain degrees of reflective spots (see Figure 1). All these problems can influence the stability of feature descriptors. Second, there're shadow and dislocation in the seam caused by panorama image stitching, and texture fusion is necessary. Besides, the conventional panorama image technology generally relies on texture projection and expansion from a square or sphere model [12]. However, the soft cavity (e.g. stomach) totally differs with standard square or sphere. If re-projected directly, it will result in huge distortion. Therefore, it's necessary to develop a new texture projecting model.

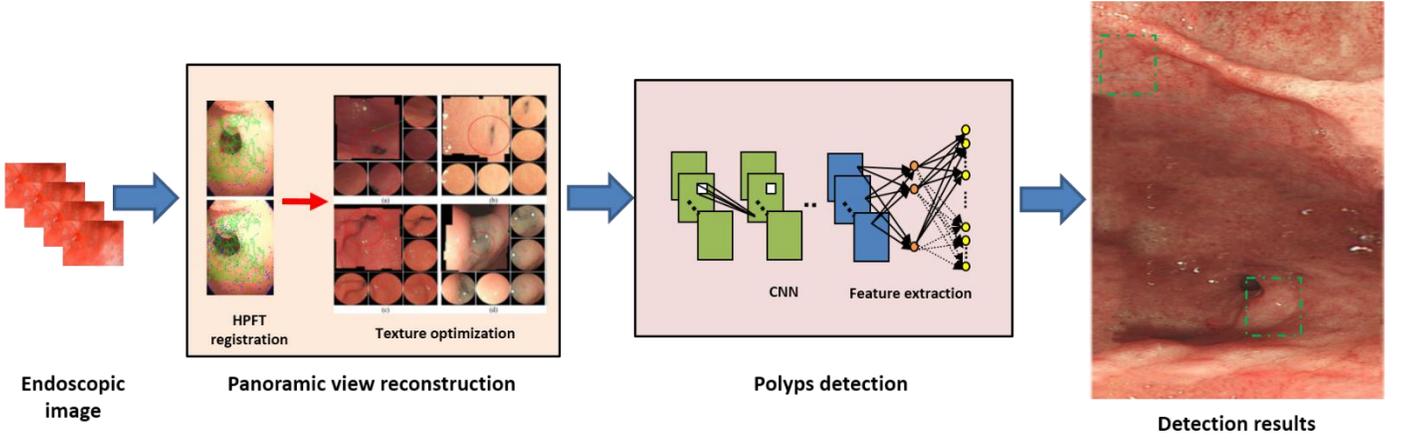

FIGURE 2: The pipeline of our method. Original endoscopic images are used to generate panoramic result. Then polyps are detected with our deep learning framework.

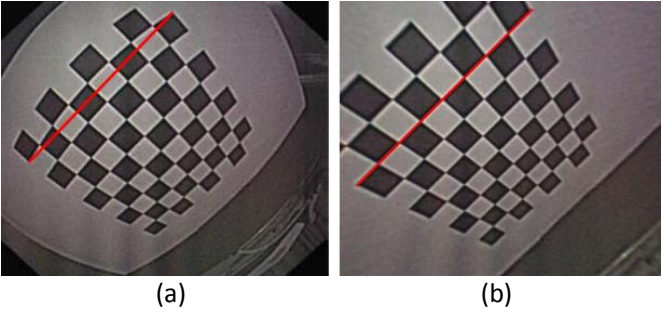

(a)           (b)

FIGURE 3: (a) is originally captured by endoscope, the chessboard is badly distorted (red line can be considered as reference). (b) is calibrated results.

## 2. Panoramic View Construction
### 2.1. Image Registration

In previous work, researchers used electromagnetic tracking device to estimate real-time coordinates of endoscope in the stomach to solve the matching problem [7]. However, this method can't be applied in traditional endoscope examination, without hardware improving. In this paper, for the texture feature of soft organs, we adopt the Homographic Patch Feature Transform (HPFT) based on homology hypothesis [13]. The public data shows that relative to the other patch feature transform of traditional computer vision, HPFT is more efficient in scene of soft organs represented by stomach, of which the main motion feature is regional peristalsis.

First of all, we give an overview of HPFT. HPFT uses some patch feature transform to detect initial feature points, such as SIFT. If local image blocks of image registration satisfy the homographic principle of computer vision, as.

$$m' = \rho H m \quad (1)$$

In the formula (1), $m'$ and m represent image local feature point pair to be determined. H represents homographic relationship that can be shown as 3*3 matrix. ρ represents scaling scale. On basis of the homographic principle, it divides image sub-blocks for initial matching feature points. By means of KL similarity, it verifies the similarity of image sub-blocks. It iteratively subdivide for the inner heard of the image sub-block, until the subdivided image sub-blocks satisfy the homographic hypothesis, for the sub-blocks that don't satisfy the homographic hypothesis.

Our experimental result shows we get a better-distributed matching feature points by adopting HPFT, compared to patch feature transform of computer vision as SIFT, FAST and so on (see Figure 4).

The other problem during image registration is to exclude the mismatching image feature point pairs. Conventional method adopts external polar line constraints and other methods to filter the matching results of image feature points. However, the accuracy of external polar line highly relies on camera viewing. So the filtering result is not satisfactory. In this paper, we regard the entire process from gastroscope entering into stomach to leaving as a closed chain process of image registration, and use closed chain optimization to filter and exclude the mismatching points of the image registration.

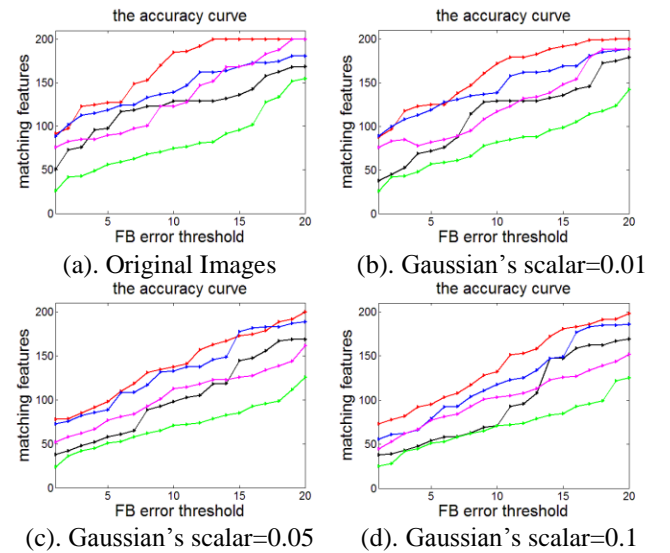

(a). Original Images    (b). Gaussian's scalar=0.01
(c). Gaussian's scalar=0.05    (d). Gaussian's scalar=0.1

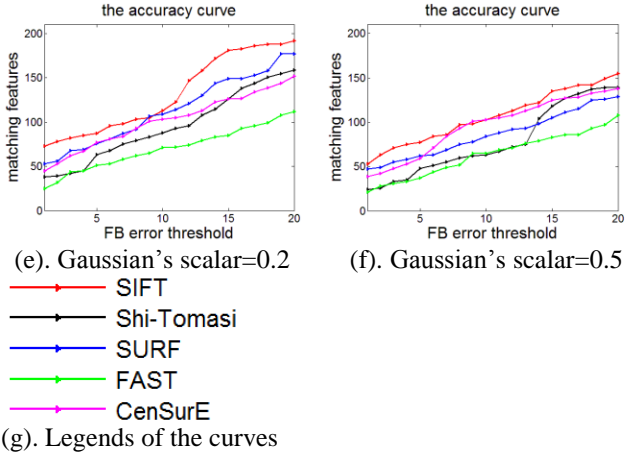

(e). Gaussian's scalar=0.2  (f). Gaussian's scalar=0.5

— SIFT
— Shi-Tomasi
— SURF
— FAST
— CenSurE

(g). Legends of the curves

FIGURE 4: Five tested method registration results. From (a) to (f), the registration methods were applied to original gastroscopic and noise images (the Gaussian noise scalar varies from 0.01 to 0.5). For each registration method, the initial detected feature number was 200, thus the ideal matching features' number was also 200. In the figures, the color curves represent the number of the detected features whose FB error is smaller than the corresponding FB error threshold (unit: pixels). The figure is quoted from [13] and [24].

2.2. **Optical Consistency Texture Fusion**.

After the image registration, the major problem in building panorama image is how to deal with stitching problem between the borders of the stitched image blocks after directly using the result of image registration. The stitching problem is generally caused by the two following reasons: first, in the result of registration obtained by HPFT and closed chain optimization, there are mismatching point pairs exists inevitably, which lead to the error of transformation matrix between images calculated, and finally reflect on the seaming results; second, there are lots of mucosa on the gastric inner wall. The images captured from different viewings will have different degrees of glisten or even direct reflection. Even though for the images reflect a same physiological location, there may be large difference in the pixel value.

Based on above facts, we develop an optimized method of texture characterization on basis of optical consistency. This method directly regards pixel difference of minimizing stitched seam as optimized target, so as to obtain panorama image structure of gastric inner wall of smoothing process. This method will be elaborated below.

Assume the sequence of images to be matched is $I_n = \{I_1, I_2, I_3, \ldots, \}$. Stitching relationship gotten by image registration can be represented by $T_{i-1,i}$, as:

$$I_i = T_{i-1,i} \cdot I_{i-1} \quad (i-1 > 0) \qquad (2)$$

Random coordinate $(x, y)$ of the image $I_{i-1}$ converts to $(x', y')$ by T transformation. The seam pixel difference after spliced is:

$$e = \|I_{i\,(x,y)} - I_{i-1(x',y')}\| \qquad (3)$$

The above formula can be organized into an optimized objective function:

$$E_{loss}(c, T) = \sum w * e^2 + \beta T^t T \qquad (4)$$

The above formula can be seen as a function relative to pixel c and transformation matrix T. In this paper, so as to simplify the process, we regard c as gray average value of the corresponding pixel on the image. Besides, ω represents the weight of value between the pixel differences, and $\beta T^i T$ represents optimized regular term, so as to avoid over fitting.

We use iterative method to optimize the formula (4). Firstly, assuming that c is unchanged, we can regard image as a four-degree-of-freedom vector $T_k = (r_1, r_2, t_1, t_2)$, $r_1, r_2$ represent rotating information. $t_1, t_2$ represent moving information. On basis of above assume, formula (4) can be seen as a linear optimization equation, and then use KL or Gauss Newton method to solve the equation. When T is fixed, the formula (4) is equivalent to solving the average of pixel values spliced to some coordinate.

For texture projection model, we adopt double cube projection model to project and expend the texture [18, 19]. Relative to conventional cube or sphere projection method, the double cube projection model can do better to deal with deformation problem after model expending.

## 3. Automatic Detection for polyps

After building the panorama image of the gastric inner wall, we develop an automatic detection technology for polyps to assist doctors in diagnosis. In the field of computer vision, this type of issues belongs to object detection. Currently, the mainstream object detection technology adopts deep learning frame to recur and fit the target region labeled by images, such as faster RCNN [20] and SSD [21]. In the scene of panorama image, the problems we need to solve are: first, how to collect large amount of manual data labeled by panorama image and second, traditional object detection based on deep learning mostly consider natural images as input. Currently, there aren't publicly published technical solutions based on panoramic images.

### 3.1 Network Model

First of all, we need to develop the network model for gastroscopic panorama image detection. In the literatures of deep learning, the classic idea of object detection technology is like this: it firstly predicts potential local bounding box on the image, and then determines the category prediction confidence of bounding box in the thought of image classification. Finally, via a series of post-processing optimization, the final results are finetuned. The representative technology of this method is RCNN and Faster RCNN [20], which is the most universal technology in the scene of target detection. However, the over-complicated frame cascade structure and the slow convergence and performance are the existing problem. In this paper, we adopt SSD [21] network frame to achieve the target detection of panorama image. Relative to RCNN and other related methods, SSD technology converts target bounding box detection, category prediction and bounding box optimization to paralleled CNN convolution. Relative to Faster RCNN and other related methods, SSD technology has the advantage of faster convolution and more accurate in target locating [22].

During the panorama data processing, because the polyps are relatively small in the panorama image, after SSD prediction most of bounding boxes are negative samples. Compared to predict directly the original images, a large amount of negative samples will lead to the unbalance of training sample sets and be hard to converge in training. For these issues, we propose an optimized SSD called selective SSD. In selective SSD, we sort negative samples detected in the process of iteration on SSD original loss function, and only select the negative samples with high confidence for training, directly filtering the low ones. This method solves the network convolution problem, and improves the accuracy of convergence to a certain degree (see Table 4).

**3.2 Sample Collection**

In addition to network model, to get the training data of panorama image is also a key in polyps' detection. In this paper, based on gastroscopic examination system of Zhejiang University Affiliated Hospital, we recruit clinical experts to mark the polyps in original gastric images. Multiple-check solution is adopted. Only the polyps these are found by two more experts can be considered as true positive samples. Then we use the proposed panorama technology to construct panorama image and the global truth is marked by clinical experts. During the data training, apart from clinical experts' marking, we adopt data enhancement technology to obtain more training samples. The enhancement method is like this: for any panorama image originally captured, we take advantage of Gaussian function for local smoothing with different scales.

## 4. Experimental evaluation and results

So as to verify the accuracy of polyps detection method based on panorama image that we proposed in this paper, we embed the developed system into gastroscopy system of the gastroenterology in Zhejiang University Affiliated Hospital. The endoscope device is GIF-QX-420 from Olympus, Japan. The frame rate of the images collected by this endoscope is 30 Frame per Second. And the image resolution is 560*480. All the volunteers recruited have a history of moderate or severe gastrointestinal diseases, 43 cases in total. Patients provide written informed consent. The clinical data collected can be used for evolution and follow-up visits. During the experiment, there is not any adverse event. Volunteer information is described in Table 1.

Gastroenterologists can construct panorama image without extra interoperations. During the examination, this system extracts HPFT feature descriptors in real time, on basis of gastric inner wall texture information collected by doctors. Then, it estimates real-time position of camera. The panorama image is gradually constructed and unfolded, as well as the polyps in panorama image are detected in real time, shown in Figure 5.

TABLE 1. Volunteer Information

|  | Accuracy Percent |
|---|---|
| Average age (range) | 54(40-64) |
| male | 32(75%) |
| smoke | 27(65%) |
| alcohol | 39(90%) |
| Intestinal metaplasia(mild/moderate/severe) | 0/12/31 |

**4.1 Panorama Assessment**

First we evaluate the performance of the panoramic result. Accordingly, our method is easier compared with the ones that construct gastroscopic panorama image based on electromagnetic guided tracking device [19]. This paper adopts the proposed texture metric error [16] to estimate the accuracy of panorama image, and the results are demonstrated in Table 2. Generally speaking, the average error of the results is 0.33, and the effect is better than the published papers [19].

TABLE 2. Quantitative Evaluation about Panorama Results (43 volunteers). We evaluate the error score between Liu's [19] method and our method. The overall texture error of ours is 0.33, which is much better than [19]. Moreover, we also evaluate the results on angularis, antrum, and stomach body respectively, and our method is better.

|  | Texture Metric Error | | | |
|---|---|---|---|---|
|  | angularis | antrum | stomach body | overall |
| Liu's method[19] | 0.53 | 0.49 | 0.44 | 0.49 |
| Our method | 0.38 | 0.35 | 0.27 | 0.33 |

## 4.2 Polyp Detection Evaluation

To evaluate the accuracy of the polyp detection, we compare the results marked by clinical doctors to the ones of automatic detection for polyps (see table 3). We consider the IOU >0.5 as positive samples, and IOU indicates that the intersection area between the area clinical doctors marks and the detection area that our method generates.

Finally the comparing results show in the case of a recalling rate close to 100%, the accuracy is 95%, which meets the requirements of clinical auxiliary diagnosis.

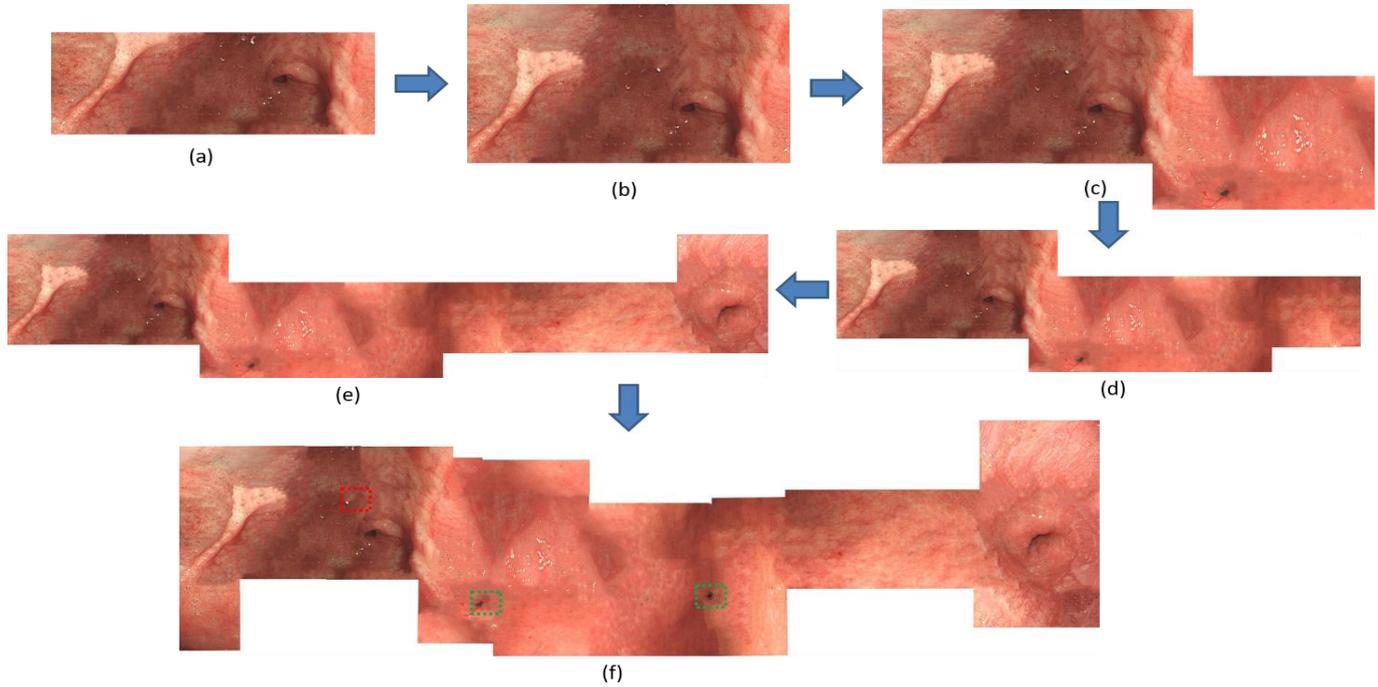

FIGURE 5: the experimental result.

TABLE 3. Polyp Detection Compared with Clinical Diagnosis. We evaluate the recall percentage and accuracy from different physiological location

|  | Clinical Diagnosis | Our Method | Recall Percentage | IOU>0.5 | IOU<=0.5 | Accuracy |
|---|---|---|---|---|---|---|
| angularis | 58 | 58 | 100% | 2 | 56 | 96.5% |
| antrum | 71 | 71 | 100% | 4 | 67 | 94.4% |
| stomach body | 40 | 40 | 100% | 1 | 39 | 97.5% |

Furthermore, we develop the selective SSD object detection framework for dealing with panoramic images, as a result, we also evaluate the performance of the selective SSD and other object detection method.

TABLE 4. Different Deep Learning Framework for Polyp Detection.

|  | Accuracy | Recall |
|---|---|---|
| Selective SSD | 95% | 100% |
| Original SSD | 84% | 75% |
| Faster RCNN | 81% | 84% |
| RCNN | 79% | 71% |

In table 4, we can see that the selective SSD outperforms over the other published method. And there is a significant improvement between selective SSD and original SSD, which indicates prove that our method is effective.

## Discussion and Conclusion

Gastroscopy is one of the most routine solutions in the current gastric diseases. It is of importance to improve the efficiency of diagnosis and reduce the risk of misdiagnosis clinically. Compared with published papers, we first put forward an end-to-end full-view automatic detection technology. Specifically, we research a method to assist doctors in understanding the whole picture of human stomach by means of panorama gotten without extra device. Then on the basis of panorama image, we propose a polyp auxiliary diagnosis method based on deep learning framework, which can improve the efficiency of doctors' diagnosis. The method is a good reference for endoscope intervention diagnosis in other human soft organs theoretically.

## Conflict of Interests

The authors declare that there are no conflicts of interest regarding the publication of this paper.

## Acknowledgments

The project was supported by the Medical Scientific Research Foundation of Zhejiang Province, China (2018PY022）.